\def\verPreprint{1}
\def\verPAPER{2}
\def\ver{1}

\ifx\ver\verPAPER
\fi

\ifx\ver\verPreprint
\documentclass[a4paper,english,11pt]{article}
\usepackage[bindingoffset=0.5cm,left=2.5cm,right=2.5cm,top=2.5cm,bottom=2.5cm,footskip=1.0cm]{geometry}
\usepackage{authblk}
\fi
\ifx\ver\verPAPER
\RequirePackage{fix-cm}
\documentclass[preprint,12pt]{elsarticle}
\newcounter{bla}

\fi

\usepackage{amsmath}
\usepackage{amssymb}
\usepackage[english]{babel}
\usepackage{dcolumn}
\newcolumntype{d}[1]{D{.}{.}{#1}}
\usepackage{float}
\usepackage{graphicx}
\usepackage{hyperref}
\usepackage{lineno}
\usepackage{multirow}
\usepackage{xcolor}
\usepackage{xspace}

\ifx\ver\verPAPER
\modulolinenumbers[5]
\fi

\newcommand{\Pagne}{\ensuremath{\bar{\nu}_{\textrm{e}}}\xspace}
\newcommand{\Pagngm}{\ensuremath{\bar{\nu}_{\mu}}\xspace}

\newcommand{\Pa}{\ensuremath{\textrm{a}_{1}}\xspace}
\newcommand{\Pai}{\ensuremath{\textrm{a}_{1}\textrm{(1260)}}\xspace}

\newcommand{\Pem}{\ensuremath{\textrm{e}^{-}}\xspace}

\newcommand{\Pfii}{\ensuremath{\textrm{f}_{2}\textrm{(1270)}}\xspace}

\newcommand{\Pggx}{\ensuremath{\gamma^{\ast}}\xspace}

\newcommand{\Pgmm}{\ensuremath{\mu}^{-}\xspace}

\newcommand{\Pgngt}{\ensuremath{\nu_{\tau}}\xspace}
\newcommand{\Pgp}{\ensuremath{\pi\xspace}}
\newcommand{\Pgpm}{\ensuremath{\pi^{-}}\xspace}
\newcommand{\Pgpp}{\ensuremath{\pi^{+}}\xspace}

\newcommand{\Pgpz}{\ensuremath{\pi^{\textrm{0}}}\xspace}
\newcommand{\Pgr}{\ensuremath{\rho}\xspace}
\newcommand{\Pgra}{\ensuremath{\rho\textrm{(1450)}}\xspace}

\newcommand{\Pgt}{\ensuremath{\tau}\xspace}
\newcommand{\Pgtm}{\ensuremath{\tau^{-}}\xspace}
\newcommand{\Pgtp}{\ensuremath{\tau^{+}}\xspace}

\newcommand{\PH}{\ensuremath{\textrm{H}}\xspace}

\newcommand{\PZ}{\ensuremath{\textrm{Z}}\xspace}
\newcommand{\h}{\ensuremath{\textrm{h}}\xspace}
\newcommand{\s}{\ensuremath{\textrm{s}}\xspace}
\newcommand{\mtau}{\ensuremath{m_{\Pgt}}\xspace}
\newcommand{\gammava}{\ensuremath{\gamma_{\textrm{va}}}\xspace}
\newcommand{\J}{\ensuremath{\textrm{J}}\xspace}
\newcommand{\Jstar}{\ensuremath{\J^{\ast}}\xspace}
\renewcommand{\P}{\ensuremath{\textrm{P}}\xspace}
\newcommand{\N}{\ensuremath{\textrm{N}}\xspace}
\newcommand{\GeV}{\ensuremath{\textrm{GeV}}\xspace}

\renewcommand{\L}{\ensuremath{\textrm{L}}\xspace}
\newcommand{\cf}{cf.\xspace}

\ifx\ver\verPAPER
\journal{Computer Physics Communications}
\fi

\begin{document}

\ifx\ver\verPAPER
\begin{frontmatter}
\fi

\title{The polarimeter vector for $\Pgt \to 3 \Pgp\Pgngt$ decays}
\ifx\ver\verPreprint
\author[1]{Vladimir Cherepanov}
\author[2]{Christian Veelken}
\affil[1]{University of Florida, Department of Physics,  Gainesville, FL, 32611 - 8440, USA }
\affil[2]{National Institute of Chemical Physics and Biophysics (NICPB), R\"{a}vala pst 10, 10143 Tallinn, Estonia}
\fi
\ifx\ver\verPAPER
\author[a]{Vladimir Cherepanov\corref{author}}
\author[b]{Christian Veelken}
\cortext[author] {Corresponding author.\\\textit{E-mail address:} vladimir.cherepanov@cern.ch}
\address[a]{University of Florida, Department of Physics,  Gainesville, FL, 32611 - 8440, USA }
\address[b]{National Institute of Chemical Physics and Biophysics (NICPB), R\"{a}vala pst 10, 10143 Tallinn, Estonia}
\fi

\ifx\ver\verPreprint
\maketitle
\fi

\begin{abstract}
The polarimeter vector of the $\Pgt$ represents an optimal observable for the measurement of the $\Pgt$ spin. 
In this paper we present an algorithm for the computation of the $\Pgt$ polarimeter vector for the decay channels $\Pgtm \to \Pgpm\Pgpp\Pgpm\Pgngt$ and $\Pgtm \to \Pgpm\Pgpz\Pgpz\Pgngt$.
The algorithm is based on a model for the hadronic current in these decay channels, which was fitted to data recorded by the CLEO experiment
\cite{CLEO:1999rzk}.
\end{abstract}

\ifx\ver\verPAPER
\begin{keyword}
Tau; Polarimeter vector
\end{keyword}
\end{frontmatter}
\fi


\ifx\ver\verPAPER
{\bf PROGRAM SUMMARY} \\
\begin{small}
\noindent
{\em Program Title: PolarimetricVectorTau2a1, version 1.0.1} \\
{\em CPC Library link to program files:} (to be added by Technical Editor) \\
{\em Developer's repository link:} https://github.com/TTauSpin/PolarimetricVectorTau2a1 \\
{\em Code Ocean capsule:} (to be added by Technical Editor)\\
{\em Licensing provisions(please choose one):} 
MIT \\
{\em Programming language: C++ 11} \\
{\em Nature of problem(approx. 50-250 words):}\\
  The polarimeter vector $\h$ of the $\Pgt$ can be used to measure the $\Pgt$ spin orientation.
  The vector $\h$ is a function of the momenta of the particles produced in the $\Pgt$ decay and needs to be computed in the restframe of the $\Pgt$ lepton. 
  While for the decay channels $\Pgtm \to \Pgpm\Pgngt$ and $\Pgtm \to \Pgpm\Pgpz\Pgngt$ expressions for $\h$ exist in the literature,
  no corresponding expressions exist for the channels $\Pgtm \to \Pgpm\Pgpp\Pgpm\Pgngt$ and $\Pgtm \to \Pgpm\Pgpz\Pgpz\Pgngt$. \\
{\em Solution method(approx. 50-250 words):}\\
  In this paper, we present an algorithm for the computation of the $\Pgt$ polarimeter vector $\h$ for the decay channels $\Pgtm \to \Pgpm\Pgpp\Pgpm\Pgngt$ and $\Pgtm \to \Pgpm\Pgpz\Pgpz\Pgngt$.
  The algorithm is based on a model for the dynamics of hadronic interactions in these decay channels.
  The parameters of the model have been determined by a fit to data recorded by the CLEO experiment.
   \\
\end{small}   
\fi

\section{Introduction}
\label{sec:Introduction}

The polarimeter vector $\h$ of the $\Pgt$ lepton represents an optimal observable for the measurement of the $\Pgt$ spin orientation $\s$.
Measurements of $\s$ allow the determination of the Weinberg angle~\cite{ALEPH:2001uca,DELPHI:1999yne,L3:1998oan,OPAL:2001brm,ATLAS:2017xuc,CMS:2023mgq} and of the $CP$ properties of the Higgs boson~\cite{ATLAS:2022akr,CMS:2021sdq}.
The term {\em optimal} refers to the fact that $\h$ provides equal or higher sensitivity for the measurement of $\s$ compared to alternative observables~\cite{Davier:1992nw}.
The polarimeter vector is a function of the momenta of the particles produced in the $\Pgt$ decay and needs to be computed in the restframe of the $\Pgt$ lepton. 

While the charged decay products of the $\Pgt$ are typically reconstructed with high efficiency and purity, the reconstruction of the neutral decay products is experimentally more challenging. The reconstruction of the hadronic $\Pgt$ decay mode further requires the classification of individual charged and neutral particles as either resulting from the hadronic $\Pgt$ decay or not, which constitutes a non-trivial task, in particular at the LHC. See Refs.~\cite{Behnke:2013lya,Tran:2015nxa,ATLAS:2015boj,Xu:2017lgs,CMS-DP-2020-041,CMS:2022prd,Giagu:2022gmq,ATLAS:2022aip} for recent results on the $\Pgt$ decay mode reconstruction. The computation of the $\Pgt$ lepton restframe requires the reconstruction of the neutrino produced in the $\Pgt$ decay. The use of energy and momentum conservation allows for a direct reconstruction of the neutrino momentum in electron--positron collisions~\cite{Altakach:2022ywa,Ehataht:2023zzt}, while more sophisticated methods are in use at the LHC~\cite{Elagin:2010aw,Bianchini:2016yrt,Sauerland:1358627,Cherepanov:2018npf}.

It has been demonstrated~\cite{Kuhn:1995nn} that all hadronic decay channels of the $\Pgt$ lepton provide the same optimal sensitivity, referred to as {\em $\Pgt$ spin analyzing power}, provided that the momentum of the $\Pgt$ lepton is known and all mesons produced in the $\Pgt$ decay are reconstructed and measured with negligible experimental resolution. The spin analyzing power of leptonic $\Pgt$ decays is limited to about $40\%$ relative to the sensitivity of hadronic $\Pgt$ decays~\cite{Davier:1992nw}.

Expressions for the $\Pgt$ polarimeter vector for the decay channels $\Pgtm \to \Pgpm\Pgngt$ and $\Pgtm \to \Pgpm\Pgpz\Pgngt$ are given in Ref.~\cite{Jadach:1990mz}. 
In this paper, we describe an algorithm for computing the polarimeter vector for the decay channels $\Pgtm \to \Pgpm\Pgpp\Pgpm\Pgngt$ and $\Pgtm \to \Pgpm\Pgpz\Pgpz\Pgngt$. Both decay channels proceed predominantly through an intermediate $\Pai$ resonance and are related by isospin symmetry.
We collectively refer to both channels as $\Pgt \to 3 \Pgp\Pgngt$ decays.
Corresponding expressions for $\Pgtp$ decays can be obtained via charge conjugation.
The computation of $\h$ proceeds in two steps. The first step is specific to the decay channels $\Pgtm \to \Pgpm\Pgpp\Pgpm\Pgngt$ and $\Pgtm \to \Pgpm\Pgpz\Pgpz\Pgngt$. It pertains to the computation of the hadronic current and is detailed in Section~\ref{sec:HadronicCurrent}.
The second step is identical for all hadronic $\Pgt$ decays and is described in Section~\ref{sec:PolarimeterVector}.
We conclude the paper with a summary in Section~\ref{sec:Summary}.
\ifx\ver\verPreprint
The code of the algorithm presented in this paper is available at Ref.~\cite{christian_veelken_2023_8113344}.
\fi

\section{Computation of the hadronic current}
\label{sec:HadronicCurrent}

The hadronic current, denoted by the symbol $J$, encapsulates the dynamics of hadronic interactions that are specific to each $\Pgt$ decay channel.
For $\Pgt \to 3 \Pgp\Pgngt$ decays,
it is not possible to derive an analytic expression for $J$ based on theory.
Instead, one needs to assume a model and then determine the parameters of the model by a fit to experimental data.
Different models for the decay $\Pgt \to 3 \Pgp\Pgngt$ have been proposed~\cite{Bowler:1988kf,Isgur:1988vm,Kuhn:1990ad,Feindt:1990ev,Kuhn:1992nz,GomezDumm:2003ku,Dumm:2009va} and fitted to data~\cite{ARGUS:1992olh,OPAL:1997was,DELPHI:1998bhv,Shekhovtsova:2012ra,Nugent:2013hxa}.
Our algorithm is based on the fit performed by the CLEO collaboration to data recorded in the decay channel $\Pgtm \to \Pgpm\Pgpz\Pgpz\Pgngt$~\cite{CLEO:1999rzk}.
The CLEO fit and the fit based on the Resonance Chiral theory~\cite{GomezDumm:2003ku,Dumm:2009va,Ecker:1988te,Ecker:1989yg} are implemented in recent versions of the Monte Carlo (MC) program \textsc{TAUOLA}~\cite{Jadach:1990mz,Shekhovtsova:2012ra,Nugent:2013hxa}. See Ref.~\cite{Was:2015laa} for a comparison of both fits. The CLEO fit is also used to compute the hadronic current for the decay channels $\Pgtm \to \Pgpm\Pgpp\Pgpm\Pgngt$ and $\Pgtm \to \Pgpm\Pgpz\Pgpz\Pgngt$ in the program TauSpinner~\cite{Czyczula:2012ny}. The latter allows to simulate different $\Pgt$ polarization and spin correlation states via MC reweighting.

The CLEO fit includes seven resonances to model the substructure of the $\Pai$ decay and makes the following ansatz for the contribution of these resonances to the hadronic current:
\begin{equation}
\J^{\mu} = B_{\Pa}(s) \, \sum_{i} \, \beta_{i} \, j_{i}^{\mu} \, ,
\label{eq:J}
\end{equation}
where the symbol $B_{\Pa}(s)$ denotes the Breit-Wigner function of the $\Pai$ meson, $\beta_{i}$ are complex coupling constants, and the symbols $j_{i}^{\mu}$ refer to the amplitudes of the seven resonances.

The Breit-Wigner function $B_{\Pa}(s)$ of the $\Pai$ meson in Eq.(~\ref{eq:J}) is given by:
\begin{equation}
B_{\Pa}(s) = \frac{1}{s - m_{0\Pa}^{2} + i \, m_{0\Pa}^{2} \Gamma_{\textrm{tot}}^{\Pa}(s)} \, ,
\label{eq:Ba1}
\end{equation}
where $s = a^{2}$ corresponds to the square of the center-of-mass energy of the three-pion system.
The symbol $a$ denotes the four-momentum of the $3 \Pgp$ system: $a = p_{1} + p_{2} + p_{3}$,
where $p_{1}$ and $p_{2}$ refer to the two pions of same charge and $p_{3}$ refers to the pion of opposite charge.
We use $m_{0\Pa}^{2} = 1.331$~\GeV, given in Table VI of Ref.~\cite{CLEO:1999rzk}, for the mass of the $\Pai$ resonance.

The width $\Gamma_{\textrm{tot}}^{\Pa}(s)$ increases with $s$ as new decay channels open up.
We have digitized the graph for $\Gamma_{\textrm{tot}}^{\Pa}(s)$ shown in Fig.~9 (b) of Ref.~\cite{CLEO:1999rzk} and interpolate linearly between the digitized values.
Up to an overall sign, which cancels when computing the polarimeter vector $h$, the hadronic current is symmetric with respect to the interchange $p_{1} \leftrightarrow p_{2}$.

The complex coupling constants $\beta_{i}$ as well as the masses and widths of the seven resonances are parameters of the fit.
The values determined by the fit to CLEO data for the decay channel $\Pgtm \to \Pgpm\Pgpz\Pgpz\Pgngt$ are given in Tables~\ref{tab:beta} and ~\ref{tab:mass}.
The symbol $\L$ denotes the angular momentum of the resonances, where $\L=0$, $1$, and $2$ for the $s$-, $p$-, and $d$-wave contributions, respectively.

\begin{table}[ht!]
\centering
\begin{tabular}{lcd{6.2}d{4.0}}
\multicolumn{1}{l}{Resonance} & \multicolumn{1}{c}{$\L$} & \multicolumn{1}{c}{$\vert\beta_{i}\vert$} & \multicolumn{1}{c}{$\phi_{i}/\pi$} \\
\hline
$\Pgr$                          & $s$-wave & $1.00$ & $ 0.00$ \\
$\Pgra$                         & $s$-wave & $0.12$ & $ 0.99$ \\
$\Pgr$                          & $d$-wave & $0.37$ & $-0.15$ \\
$\Pgra$                         & $d$-wave & $0.87$ & $ 0.53$ \\
$\Pfii$                         & $p$-wave & $0.71$ & $ 0.56$ \\
$\sigma$                        & $p$-wave & $2.10$ & $ 0.23$ \\
$\textrm{f}_{0}\textrm{(1370)}$ & $p$-wave & $0.77$ & $-0.54$ \\
\end{tabular}
\caption{
  Moduli $\vert\beta_{i}\vert$ and phases $\phi_{i}$ of the complex coupling constants $\beta_{i}$ (values reproduced from Table III of Ref.~\cite{CLEO:1999rzk}).
}
\label{tab:beta}
\end{table}

\begin{table}[ht!]
\centering
\begin{tabular}{ld{5.1}d{5.1}}
\multirow{2}{18mm}{$\textrm{Y}$} & \multicolumn{1}{c}{$m_{0}^{\textrm{Y}}\,\,$} & \multicolumn{1}{c}{$\Gamma_{0}^{\textrm{Y}}\,\,$} \\
 & $\,\,$[$\textrm{GeV}$] & $\,\,$[$\textrm{GeV}$] \\          
\hline
$\Pgr$                          & $0.774$ & $0.149$ \\
$\Pgra$                         & $1.370$ & $0.386$ \\
$\Pfii$                         & $1.275$ & $0.185$ \\
$\sigma$                        & $0.860$ & $0.880$ \\
$\textrm{f}_{0}\textrm{(1370)}$ & $1.186$ & $0.350$ \\
\end{tabular}
\caption{
  Masses $m_{0}^{\textrm{Y}}$ and widths $\Gamma_{0}^{\textrm{Y}}$ of resonances $\textrm{Y}$ (values reproduced from Table I of Ref.~\cite{CLEO:1999rzk}).
}
\label{tab:mass}
\end{table}

Expressions for the amplitudes $j_{i}^{\mu}$ of the seven resonances are given in Appendix A.1 of Ref.~\cite{CLEO:1999rzk}.
All form factors $F_{R_{i}}(k_{j})$ in Eq.~(A3) of Ref.~\cite{CLEO:1999rzk} are taken to be equal to one,
corresponding to the case of a point-like three-pion system (\cf Section IV.C of Ref.~\cite{CLEO:1999rzk}).
The expressions read:
\begin{eqnarray}
j_{1}^{\mu} & = & T^{\mu\nu} \, \left( B_{\Pgr}^{P}(s_{1}) \, q_{1\nu} - B_{\Pgr}^{P}(s_{2}) \, q_{2\nu} \right) \nonumber \\
j_{2}^{\mu} & = & T^{\mu\nu} \, \left( B_{\Pgr'}^{P}(s_{1}) \, q_{1\nu} - B_{\Pgr'}^{P}(s_{2}) \, q_{2\nu} \right) \nonumber \\
j_{3}^{\mu} & = & T^{\mu\nu} \, \left( (a \cdot q_{1}) \, B_{\Pgr}^{P}(s_{1}) \, Q_{1\nu} - (a \cdot q_{2}) \, B_{\Pgr}^{P}(s_{2}) \, Q_{2\nu} \right) \nonumber \\
j_{4}^{\mu} & = & T^{\mu\nu} \, \left( (a \cdot q_{1}) \, B_{\Pgr'}^{P}(s_{1}) \, Q_{1\nu} - (a \cdot q_{2}) \, B_{\Pgr'}^{P}(s_{2}) \, Q_{2\nu} \right) \nonumber \\
j_{5}^{\mu} & = & T^{\mu\nu} \, \left( (a \cdot q_{3}) \, B_{\textrm{f}_{2}}^{D}(s_{3}) \, q_{3\nu} - \frac{1}{3} \, q_{3}^{2} \, B_{\textrm{f}_{2}}^{D}(s_{3}) \left( a_{\nu} - \frac{a \cdot h_{3}}{s_{3}} \, h_{3\nu} \right) \right) \nonumber \\
j_{6}^{\mu} & = & T^{\mu\nu} \, \left( B_{\sigma}^{S}(s_{3}) \, Q_{3\nu} \right) \nonumber \\
j_{7}^{\mu} & = & T^{\mu\nu} \, \left( B_{\textrm{f}_{0}}^{S}(s_{3}) \, Q_{3\nu} \right) \, ,
\end{eqnarray}
where 
$q_{1} = p_{2} - p_{3}$, $q_{2} = p_{3} - p_{1}$, $q_{3} = p_{1} - p_{2}$,
$h_{i} = p_{j} + p_{k}$ (with $i \not= j \not= k \not= i$),
$Q_{i} = h_{i} - p_{i}$, and $s_{i} = h_{i}^{2}$. 
The subscript $\Pgr'$ refers to the $\Pgra$ resonance.
The tensor $T^{\mu\nu}$ is defined by $T^{\mu\nu} = g^{\mu\nu} - a^{\mu}a^{\nu}/a^{2}$, where $g^{\mu\nu}$ is the metric tensor.
The symbols $B_{\textrm{Y}}^{L}(s_{i})$ denote the Breit-Wigner functions of the seven resonances.
They are functions of $s_{i}$ and of the angular momentum $\L$:
\begin{equation}
B_{\textrm{Y}}^{\L}(s_{i}) = \frac{m_{0\textrm{Y}}^{2}}{(m_{0\textrm{Y}}^{2} - s_{i}) - i \, m_{0\textrm{Y}} \, \Gamma^{\textrm{Y},\L}(s_{i})}
 \quad \mbox{with} \quad
\Gamma^{\textrm{Y},\L}(s_{i}) = \Gamma_{0}^{\textrm{Y}} \, \left(\frac{k'_{i}}{k'_{0}}\right)^{2\L+1} \, \frac{m_{0\textrm{Y}}}{\sqrt{s_{i}}} \, .
\label{eq:BY}
\end{equation}
The decay momentum $k'_{i}$ is given by:
\begin{equation}
k'_{i} = \frac{\sqrt{\left( s_{i} - (m_{j} + m_{k})^{2} \right) \, \left( s_{i} - (m_{j} - m_{k})^{2} \right)}}{2\sqrt{s_{i}}} 
  \quad \mbox{(with $i \not= j \not= k \not= i$)} \, .
\label{eq:k}
\end{equation}
The symbol $k'_{0}$ refers to the decay momentum evaluated at $s_{i} = m_{0\textrm{Y}}^{2}$.

As the decay channels $\Pgtm \to \Pgpm\Pgpz\Pgpz\Pgngt$ and $\Pgtm \to \Pgpm\Pgpp\Pgpm\Pgngt$ are related by isospin symmetry, one expects this model to be valid for the decay channel $\Pgtm \to \Pgpm\Pgpp\Pgpm\Pgngt$ also.
A subsequent CLEO analysis has confirmed this assumption~\cite{Shibata:2002uv}.

\section{Computation of the polarimeter vector}
\label{sec:PolarimeterVector}

Given the hadronic current $J$ in Eq.~(\ref{eq:J}),
the polarimeter vector $\h$ of the $\Pgt$ is computed using the relations~\cite{Jadach:1990mz,Kuhn:1982di}:
\begin{equation}
\h_{\mu} = \frac{1}{\mtau \, \omega} \, \left( \mtau^{2} \, \delta_{\mu}^{\nu} - \P_{\mu} \P^{\nu} \right) \, \left( \Pi^{5}_{\nu} - \gammava \, \Pi_{\nu}\right) \, ,
\label{eq:h}
\end{equation}
with:
\begin{equation}
\Pi_{\mu} = 2 \left( (\Jstar \cdot \N) \, \J_{\mu} + (\J \cdot \N) \, \Jstar_{\mu} - (\Jstar \cdot \J) \, \N_{\mu} \right) \, , \quad \quad \Pi^{5}_{\mu} = 2 \, \operatorname{Im} \epsilon^{\mu\nu\rho\sigma} \Jstar_{\nu} \J_{\rho} \N_{\sigma}
\label{eq:Pi}
\end{equation}
and
\begin{equation}
\omega = \P^{\mu} \left( \Pi_{\mu} - \gammava \, \Pi^{5}_{\mu} \right) \, .
\label{eq:omega}
\end{equation}
In these formulas, $\mtau$ denotes the mass of the $\Pgt$ lepton,
$\P$ the $\Pgt$ lepton momentum, and $\N$ the momentum of the $\Pgt$ neutrino. The latter can be computed by taking the difference between $\P$ and the momenta of the charged and neutral mesons produced in the $\Pgt$ decay.
The ratio of the vector and axial-vector couplings of the $\Pgt$ lepton, $\gammava$, is assumed to have the Standard Model value $\gammava = 1$.
The relations given by Eqs.~(\ref{eq:h}) to~(\ref{eq:omega}) are valid for decays of $\Pgtm$. For decays of $\Pgtp$ the terms proportional to $\gammava$ in Eqs.~(\ref{eq:h}) and~(\ref{eq:omega}) reverse sign, \cf the text following Eq.~(3.16) in Ref.~\cite{Jadach:1990mz}.

The code for the computation of the polarimeter vector in $\Pgt \to 3 \Pgp\Pgngt$ decays
has been validated in the context of studies of the longitudinal $\Pgt$ polarization in the process $\PZ/\Pggx \to \Pgtp\Pgtm$~\cite{Cherepanov:2018wop} and of transverse spin correlations between the $\Pgtp$ and the $\Pgtm$ in the process $\PH \to \Pgtp\Pgtm$~\cite{Cherepanov:2018yqb}, using simulated proton-proton collisions at the LHC.

\section{Summary}
\label{sec:Summary}

The polarimeter vector $\h$ represents an optimal observable for measurements of the $\Pgt$ spin orientation.
Expressions for $\h$ for the decay channels $\Pgtm \to \Pgpm\Pgngt$ and $\Pgtm \to \Pgpm\Pgpz\Pgngt$ are given in Ref.~\cite{Jadach:1990mz}. In this paper, we have presented an algorithm that allows the computation of $\h$ for the decay channels $\Pgtm \to \Pgpm\Pgpp\Pgpm\Pgngt$ and $\Pgtm \to \Pgpm\Pgpz\Pgpz\Pgngt$, collectively referred to as $\Pgt \to 3 \Pgp\Pgngt$ decays. The computation is based on a model for the hadronic current $\J$, the parameters of which have been determined by a fit to data recorded in the decay channel $\Pgtm \to \Pgpm\Pgpz\Pgpz\Pgngt$ by the CLEO experiment. The model is expected to be valid for the decay channel $\Pgtm \to \Pgpm\Pgpp\Pgpm\Pgngt$ also, due to isospin symmetry.
A subsequent CLEO analysis has confirmed this assumption.
From our perspective as experimentalists at the LHC, it would be very useful if a similar fit to data in the decay channel $\Pgtm \to \Pgpm\Pgpp\Pgpm\Pgpz\Pgngt$ could be made.
The polarimeter vector for the leptonic decay channels $\Pgtm \to \Pem\Pagne\Pgngt$ and $\Pgtm \to \Pgmm\Pagngm\Pgngt$ has not been discussed in this paper. The spin analyzing power of these channels is limited to about $40\%$ relative to the sensitivity of hadronic $\Pgt$ decays.

\section*{Acknowledgements}
This work was supported by the Estonian Research Council grant PRG445.

\bibliography{PolarimetricVectorTau2a1}

\end{document}